# Scalable Peer-to-Peer Streaming for Live Entertainment Content

*E. Mykoniati, R. Landa, S. Spirou, R. Clegg, L. Latif, D. Griffin, M. Rio*

Department of Electronic and Electrical Engineering
University College London

## Abstract

We present a system for streaming live entertainment content over the Internet originating from a single source to a scalable number of consumers without resorting to centralised or provider-provisioned resources. The system creates a peer-to-peer overlay network, which attempts to optimise use of existing capacity to ensure quality of service, delivering low start-up delay and lag in playout of the live content. There are three main aspects of our solution. Firstly, a swarming mechanism that constructs an overlay topology for minimising propagation delays from the source to end consumers. Secondly, a distributed overlay anycast system that uses a location-based search algorithm for peers to quickly find the closest peers in a given stream. Finally, a novel incentives mechanism that encourages peers to donate capacity even when the user is not actively consuming content.

## 1 Introduction

Live content delivery over the Internet is highly desirable, but challenging. Removing the need for deploying specialised resources would allow anyone to broadcast and reach consumers anywhere in the world. This facility could be used by many applications – the most obvious being IPTV (Internet Protocol Television) for the distribution of live audio-visual content, but other applications such as real-time multi-user 3D environments and games could also be envisioned.

While multicast and QoS (Quality of Service) mechanisms at the network layer seem ideal for delivering live entertainment content, deployment is still limited. Lack of multicast can be overcome, although not efficiently, by using unicast connections to consumers. Lack of network layer QoS however, inevitably impacts QoE (Quality of Experience) observed as interruptions to the playback of many currently deployed client/server applications. Borrowing from traditional CDNs (Content Delivery Networks) for web content, IPTV service providers currently fence off a small portion of the Internet and deploy native multicast, traffic differentiation, abundant capacity, and dedicated infrastructure. Although this *walled garden* gives acceptable results – even for 8Mb/s HD (High Definition) H.264 TV channels – it is neither scalable nor cheap.

Application-layer, multicast-like solutions typically rely on organised distribution trees [1], so they are sensitive to *churn* (frequent peer arrivals and departures). On the other hand, file distribution systems like BitTorrent [2] achieve high resilience to churn with *swarming*: splitting the file into small data units and ensuring that these pieces are distributed amongst the set of peers participating in the download. In this way peers obtain parts of the file from many peers in parallel, increasing resilience. However, content delivered in this way can only be consumed once the full download has been completed.

Variations of BitTorrent for on-the-fly consumption of swarmed content such as PPLive [3] suffer from relatively long start-up delay and playout lag. In addition they often require provider-provisioned resources (super peers) to assist content distribution by providing supplementary upload capacity beyond the capacity provided by the consuming peers.

Our solution is a peer-to-peer live streaming system where all resources are provided by the peers themselves. All peers participate in the distribution and consumption of one or more streams produced at a single source – the *peercaster*. Upon joining a stream, peers attach themselves to other peers according to distributed topology management and capacity allocation algorithms that aim at minimising start-up delay and playout lag across the entire swarm.

The paper is organised as follows. The main aspects of the system are considered independently in Sections 2 to 6. Section 2 discusses the problems to be addressed by a swarming system delivering real-time content. Section 3 discusses the construction of overlay networks to deliver swarmed media streams with Quality of Service considerations. Section 4 discusses how the peers participating in a stream can be discovered in a scalable way. Section 5 discusses how the participation of non-consuming peers can be organised to increase the capacity of the system. Section 6 describes an incentives mechanism that enables claiming past contributions from peers with no prior interaction. Following this description of the main components of our solution, Section 7 goes on to describe how they fit together to form an overall system for streaming live media content over a fully distributed peer-to-peer network. Finally, Section 8 presents our conclusions.

## 2  Real-time Swarming

Peer-to-peer live streaming systems overcome lack of native multicast and QoS by implementing such functionality at the application layer. The resultant *overlay* receives the live stream from an *encoding element* (e.g. a capture device) at the peercaster, distributes it through the peers, and delivers it to a *decoding element* (e.g. a player) at the *consumer*. The system performance generally differs across consumers and is correlated to QoE through three metrics:

- *Start-up delay*: The time between a consumer's request to view a particular stream and the stream beginning to play. Low start-up delay allows quick switching between channels.

- *Playout lag*: The delay between a stream data unit being sent by the peercaster and the same data unit being viewed by the consumer. Low playout lag means that stream viewing is more "live".

- *Playout continuity*: the percentage of stream data units successfully played at the correct time by the consumer. High playout continuity enables viewing with minimal distortion.

Early peer-to-peer streaming systems mimicked native multicast and organised peers in a *tree* with the peercaster at the root. After an initial consumer request, the entire stream would be *pushed* to the consumer from a peer with available upload bandwidth and probably in close network proximity [1]. This scheme gives good start-up delay and playout continuity if the tree is stable under churn, and good playout lag if the tree is also short (low number of peers between peercaster and consumer). The highly dynamic Internet environment has so far kept *tree/push* systems within research settings.

Recent deployments of peer-to-peer live streaming systems [3] are based on the fundamentally different and inherently resilient *mesh/pull* (*swarming*) scheme: a stream is segmented at the peercaster into data units called *chunks* and each consumer independently optimises its playout by selecting which chunks to *pull* from which peers. The resultant interconnection graph, as shown in Figure 1, is not a regular grid nor fully connected, but it is called a mesh nonetheless. Mesh/pull streaming systems have built on the success of the BitTorrent file distribution mechanism, and owe their resilience to local optimisation of delivery and free selection of peers and chunks. Similar to BitTorrent, the set of peers exchanging chunks constitute a *swarm*. Unlike BitTorrent, chunks must leave the overlay and reach the decoder at the consumer *in order* and *in*

*real-time*. Furthermore, unlike VoD (Video on Demand), all chunks are not available at the time a peer selects the stream; rather they are created in real-time at the peercaster.

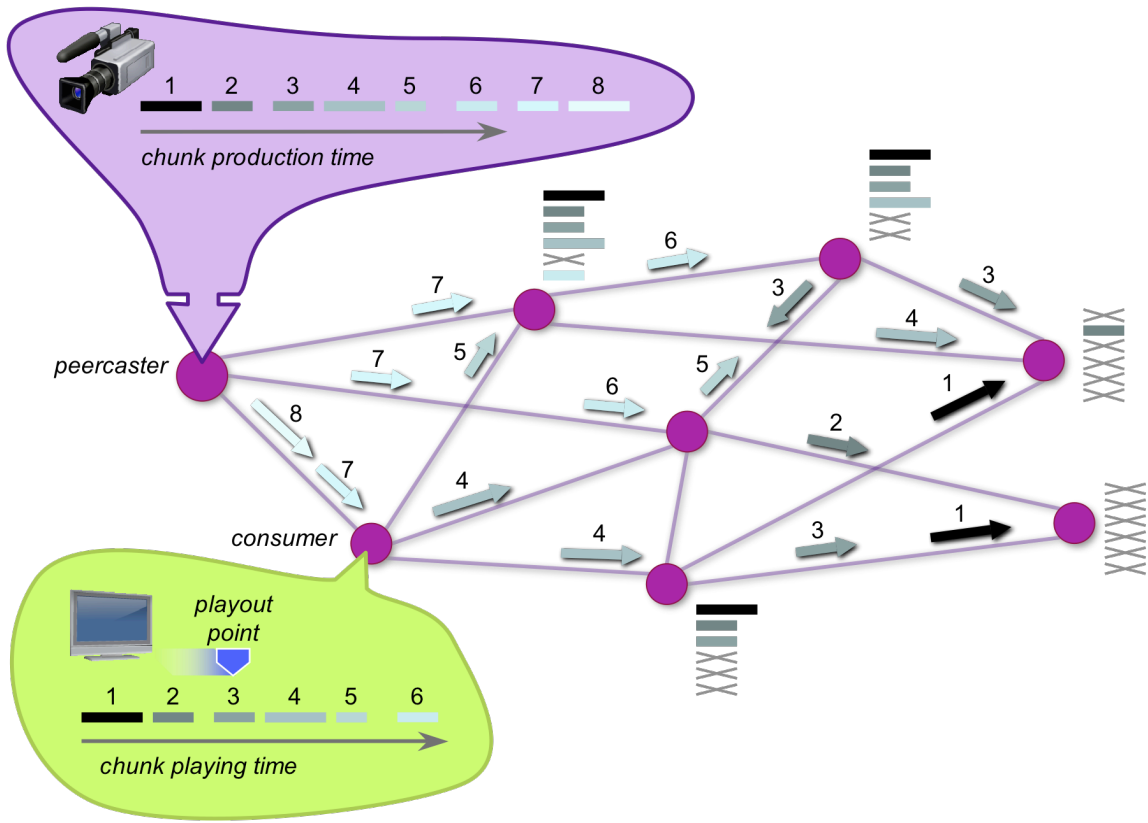

**Figure 1 – Real-time swarming**

A mesh/pull peer, requests and receives chunks from different peers in the swarm. To absorb the variance in chunk reception over different sources, and to cope with peer failures in times of high churn, a consumer maintains a *playout buffer*. The playout buffer contains chunks yet to be played out and slots for chunks yet to be retrieved. It is marked by the playout point, which indicates the next chunk to be delivered to the decoder. Playout begins once the number of received consecutive chunks is considered sufficient to support playout continuity at the stream rate, marking the start-up delay and the initial playout lag.

A large playout buffer increases playout continuity because it absorbs more jitter but it also incurs higher start-up delay and playout lag. The variance experienced in chunk reception is typically smaller for less "live" chunks, as these are already distributed across a large set of peers. More "live" chunks on the other hand are harder to find, and so retrieving them in time to preserve continuity requires a very aggressive strategy. The particular choice in the trade-off between playout lag and playout continuity depends on the capabilities and the position of each peer in the swarm, and must follow the preferences of the user, it is therefore subject to local optimisation by the peer.

The *best* performance of currently deployed swarming systems can be summarised as having 20 seconds start-up delay and 1 minute playout lag with acceptable playout continuity for streams of 350Kbit/s. This is far from even the *average* performance of current IPTV walled gardens that exhibit 2 seconds of start-up delay and 2 seconds playout lag with good playout continuity for streams of 3.5Mbit/s.

The following section discusses principles for selecting peers to retrieve chunks from, with the objective to improve both playout lag and continuity for all the peers in the system, irrespectively of their position in the trade-off between playout lag and playout continuity.

## 3  A QoS-Based Overlay

The data connections for retrieving chunks, determined by the local strategies of the individual peers, form an overlay network. The delay and throughput (QoS) over each connection and along the end-to-end path have a direct impact on the QoE experienced by each consumer. The total throughput achieved over all the incoming connections of a consumer must be enough to sustain the stream rate to ensure playout continuity. Also, minimising the playout lag is achieved by minimising the delay over the end-to-end path from the peercaster to the consumer. We define this *path delay* as the accumulated link propagation and the processing/queuing delays at each sending peer.

A poor overlay topology is one where traffic traverses long overlay links disregarding direction or locality. In addition to the negative impact on playout lag for the peers this also increases traffic in the underlying network. If the overlay is designed more intelligently, by interconnecting local peers whenever possible, this has the side-effect of benefiting the ISPs who will experience less traffic over expensive inter-domain links.

Peers need an estimate of network delay to other peers to determine which ones are in proximity. It is impractical to establish a connection, probe and measure the delay to every peer before selecting only the nearest few. To address this issue, recently developed techniques can obtain *delay space coordinates* based on only a few prior measurements [4]. Network delays are mapped to a synthetic coordinate space using metric space embedding (see Figure 2); delay between any two peers can be approximated with their geodesic metric distance in delay space.

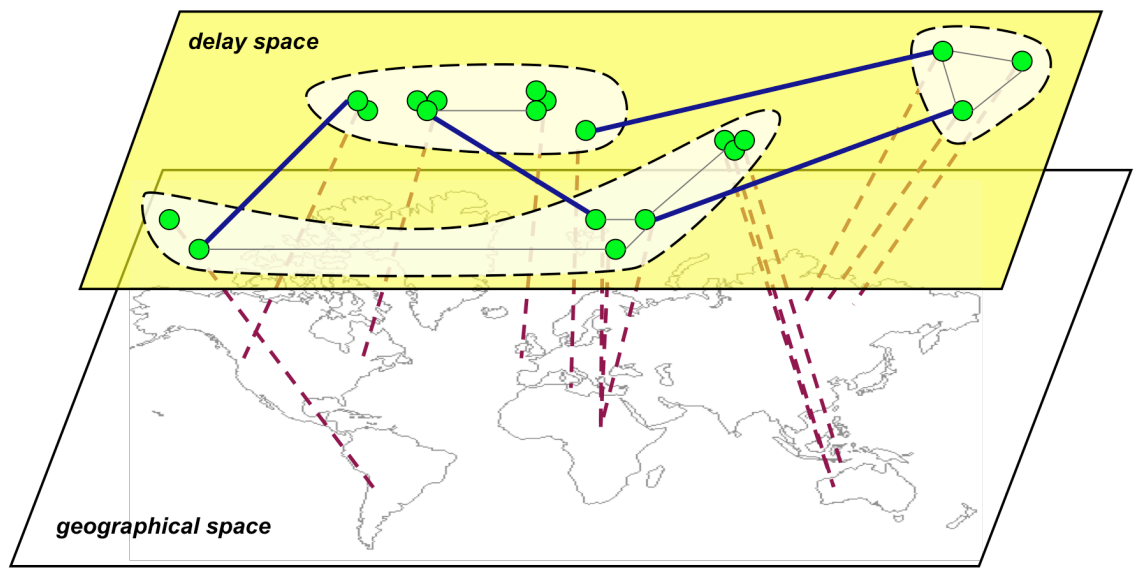

**Figure 2 – Mapping peers to synthetic delay coordinates**

A star (client/server) topology where all peers directly connect to the peercaster is the ideal for minimising delay. However, this is unrealistic as it implies that the peercaster has sufficient upload capacity to serve all peers. The best approximation of ideal direct connections to the peercaster is achieved when peers establish connections with their closest peers in the direction of the peercaster. Always favouring short connections, though, results in paths with a large hop-count, thus potentially long delays due to accumulated processing and queuing times at each hop.

Our simulation results indicate that a compromise of establishing a few longer connections to serve as "jumps" towards the peercaster is beneficial in terms of reducing hop-count without requiring increased capacity at the peercaster.

Peers are not uniformly distributed across the delay space. Rather they tend to cluster following population density but modified by the Internet topology and associated delays. Connections within the same domain have small delays, whereas one satellite inter-domain link is enough for the delay to soar. Peers that only select fast local connections may fail to reach other peers that are close enough to the peercaster to actually get the stream. Therefore, a cluster of peers needs to have enough long distance (slow) connections to reach other remote peers with the required upload capacity and access to the stream data.

Once a peer has determined the set of feasible peers it prefers on topological and delay grounds it will usually request and receive data from only a subset. Decisions on which peers to download a particular chunk from depend on the potential senders' playout lag and chunk availability and other factors such as the actual network delay experienced over the links, their available upload bandwidth, stability, trust, and past performance.

The following section elaborates on the distributed discovery of peers to establish the required connections in the direction of the peercaster.

## 4  Scalable Peer Discovery

A critical aspect of our system is the search for other peers that are available to send chunks of the stream. When first connecting to a stream a peer wishes to quickly determine which other peers can provide chunks and are also close in terms of delay. To achieve this we introduce the concept of a *local tracker* (LT). Local trackers are ordinary peers that have the additional function of listing peers carrying a stream in their *local area*. The scope of local areas is fluid: they split or merge as they become over/under-loaded as peers join/leave the stream. Figure 3 shows an example of three streams, their local areas and the underlying distribution of nodes on all streams.

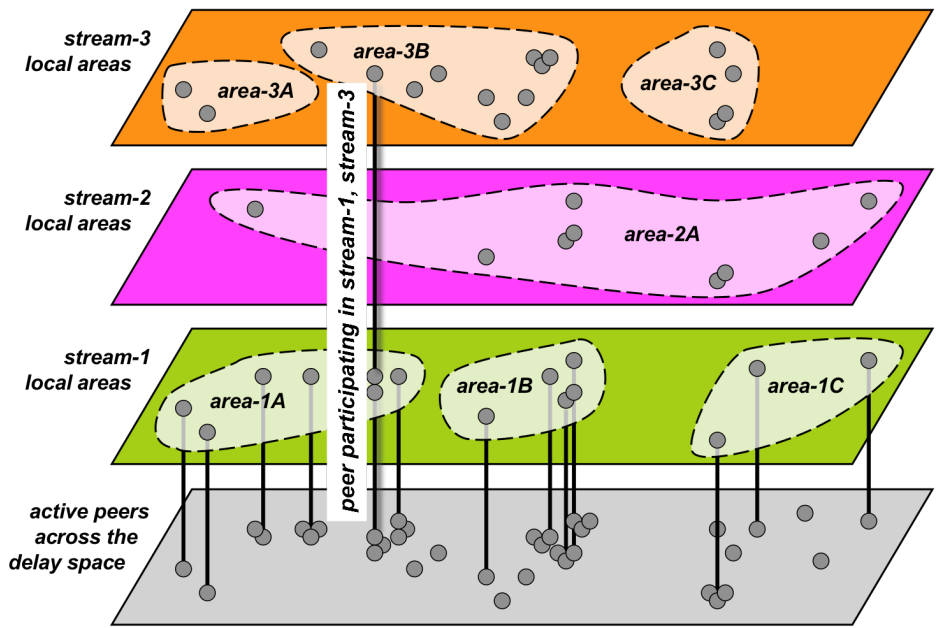

Figure 3 – Peer clustering in local areas across multiple streams

As already mentioned a low start up delay is an important goal for the system and this involves peers being able to quickly obtain a list of nearby peers carrying the stream it is interested in. To achieve this task we developed a *Distributed Overlay Anycast Table* (DOAT) inspired by the Chord DHT system [5]. The role of the DOAT is, when queried by a peer about a stream, to quickly return the address of an LT which is near to that peer in terms of delay.

To facilitate the search task within the synthetic coordinate domain, network coordinates are transformed to a single dimension. This is achieved using a space-filling curve which has the property that if two locations are "close" on the curve they are also close in the original space (but not vice versa). One particular curve that has been shown to have this property is the *H curve* [6]. Figure 4 shows a space-filling curve, generated with three iterations, and how nodes map to it. In our system, this space-filling curve replaces the ring structure of a typical Chord DHT and, thus, the distance in the DOAT ring structure corresponds to the distance in the network delay space (with the potential error introduced by the space-filling curve).

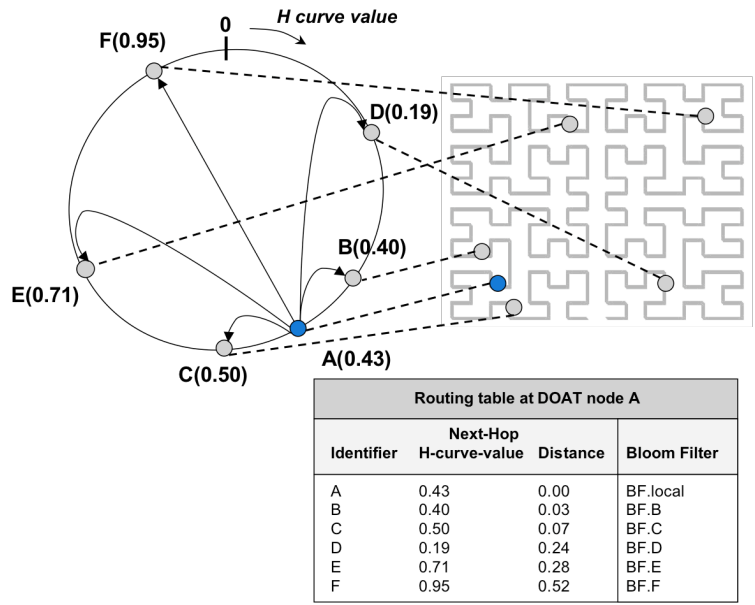

Figure 4 – Distributed Overlay Anycast Table

The routing table of a DOAT node is, similarly to Chord, populated with neighbour DOAT nodes at exponentially increasing distances in either direction of the DOAT ring. A DOAT routing entry contains the next-hop DOAT neighbour node and the list of identifiers of the streams, for which LT data can be reached through this DOAT neighbour node. To accelerate the process of matching a stream identifier to a list, and to reduce the overhead of routing table updates, the list of stream identifiers is aggregated into a *Bloom filter* [7]. In this case, the Bloom filter is the result of setting to logical 1 all the bits corresponding to each of the $k$ hash values of all the identifiers in the list, and provides an efficient data structure to store and verify set membership.

Whenever a new LT for a particular stream is added, it registers with the closest DOAT node in delay space and the stream identifier is added to the Bloom Filter of the *local* routing entry of that DOAT node. Further, if the DOAT node had no previous route for the stream identifier of the newly added LT, a routing update is sent to all its neighbours, signalling the existence of the new route. Upon receiving a routing update from a neighbour node *A*, a DOAT node *B* updates the routing entry associated with *A*, and further forwards the update to its neighbours that are further away, in terms of their distance along the space-filling curve, than node *A* (closer nodes will receive the update from nodes nearer to them). In order to reduce the frequency of messages, a minimum interval is introduced over which routing updates are aggregated and sent over a single routing update message.

To obtain the IP address of the closest LT carrying a given stream, peers issue a query with the stream identifier to their closest DOAT node. The DOAT node then searches its routing table in increasing order of distance, and the query is forwarded to the next-hop DOAT node of the first routing entry whose Bloom filter matches the stream identifier. This is done recursively at each DOAT node and the query propagates in logarithmically decreasing distances, until reaching the DOAT node with an LT for the requested stream identifier in its local routing entry. The LT discovered by this process is the closest LT to the querying peer in the requested stream.

## 5  Distributed Capacity Provisioning

Traditionally, QoS for multimedia services is provided through capacity over-provisioning and service differentiation. These are impossible for large-scale, real-time video distribution overlays with no control over their underlying infrastructure. Instead, the idea is to exploit the large, aggregate upload bandwidth of the system to serve the consuming peers. This is possible assuming that some peers will participate in the system without consuming stream content. We call these *non-consuming peers* (NCPs). Through our incentives mechanism (see Section 6), NCPs contribute to the distribution of chunks, in order to "redeem" these contributions later by "purchasing" chunks for their own consumption.

When an NCP downloads a chunk it uses the upload bandwidth of another peer. To offset this loss of system resources and to further optimise distribution an NCP conducts bandwidth multiplication by downloading only a fraction of a stream's chunks and uploading them as many times as possible. So, NCPs increase the QoS for consuming peers, instead of just distributing media traffic between themselves.

Another advantage of NCPs is the reduction of the playout lag when they are deployed appropriately. This can be accomplished by encouraging NCPs, at stream bootstrap, to join in order of their delay from the peercaster. The result is an interconnected web of bandwidth multipliers with low playout lag. Consuming peers inherit this low playout lag when they connect to a stream containing such organised NCPs.

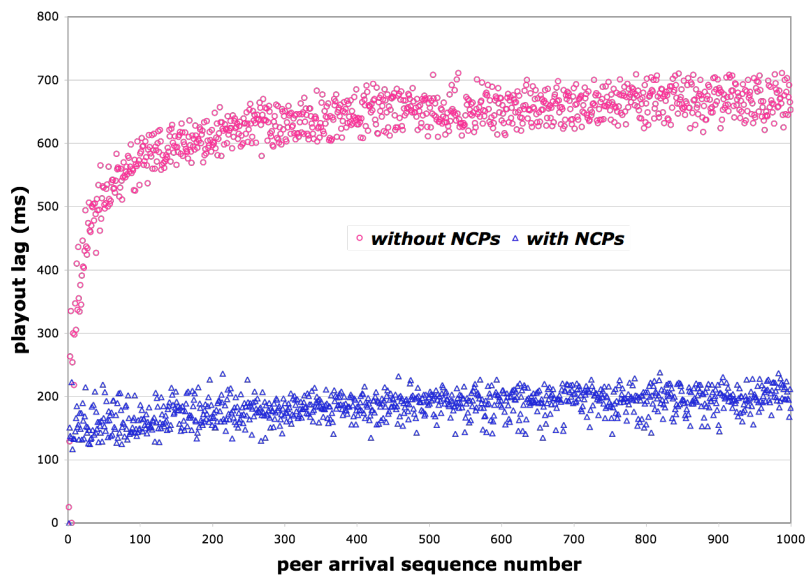

**Figure 5 – Performance enhancement with non-consuming peers**

This effect can be seen in Figure 5, depicting the playout lag for a simulation of 100 NCPs and

1000 peers, distributed uniformly across a rectangular delay space with the peercaster at the centre, and 140ms maximum delay between any peer and the peercaster.

## 6 Incentives

As it has been shown repeatedly [8], peer-to-peer systems where there is no contribution accounting and associated incentives mechanism, can suffer from pervasive freeloading and widespread quality degradation. Additionally, as detailed in Section 5, a QoS-based overlay network requires peers to contribute resources to the system even if they are not interested in consuming resources at the same time. This means that peers must be able to make contributions towards some set of peers, and then redeem those contributions at a later time, and from a different set of peers.

This problem is usually solved using reputation systems [9], where peers implement extended "word-of-mouth" recommendation networks. However, most reputation systems are vulnerable to identity-based attacks where a single peer commands legions of disposable identities, or where peers lie about the contributions of other peers in an attempt to deny them access to network resources. We propose an incentives mechanism where peers only keep track of the contributions that they have given or received from specific neighbour peers, but are potentially able to use those contributions with all peers. Additionally, our protocol is designed to ensure that peers are unable to profit from multiple identities, and do not need to rely on possibly false third-party information about the contributions of other peers.

In order to address these identity management issues peers use signed pseudonyms that their neighbours can verify with public keys exchanged when the peers initially come into contact. From then on, all communication between peers is digitally signed. Thus, contributions are always bound to an identity, and peers gain nothing from having multiple identities: their contributions will be simply split among them. Additionally, we propose the translation of contributions to an abstract numerical trust value, which can be directly manipulated by the peers and re-converted into contributions when necessary. The trust a peer $A$ has to another peer $B$, corresponds to the contribution peer $A$ has received from peer $B$ in the past, without offering anything in return, and ought therefore to return to peer $B$ in the future. Note that there is no explicit interaction for payment. Instead, each peer maintains trust from/to other peers locally.

To enable peers to contribute to a set of peers but receive contributions from another, we propose *trust shifting*. Trust shifting enables a peer to receive contribution from a peer it has no previously established trust with, by shifting trust along a path of trust relationships among other peers. Thus, any peer can perform contributions towards a given peer and then request to have this trust shifted to a different peer, from which it can now request any service.

To accomplish this, each peer constructs a local view of the *trust network*. Its immediate neighbours are found by direct experience, but neighbours further away are discovered using a truncated, self-avoiding random walk algorithm that discovers high-trust paths preferentially. In essence, each peer periodically advertises its local trust accounts. As these advertisements necessarily traverse the peers who actually performed the contributions represented in their account values, it is trivial to ensure that only truthful messages are propagated. Messages are propagated preferentially through links with high trust in a probabilistic fashion. The probability of messages being forwarded to a given neighbour peer is proportional to the local trust account of this neighbour. Thus, an announcement has greater probability of traversing paths where trust links have consistently high trust values.

In order to determine what is the maximum amount of trust that can be shifted to any particular destination along the discovered trust connections, each peer runs an instance of a MaxFlow-MinCut algorithm [10]. The shift itself is implemented through trust shift request messages that are source-routed and execute the trust exchanges in a hop-by-hop basis. Once trust has been shifted, it is essentially undistinguishable from trust acquired by direct contribution, allowing peers to flexibly decide where to "spend" their accumulated trust.

## 7 Putting it all together

As discussed in the previous sections the system consists of multiple independent swarms carrying different streams of real-time content. Peers participate in one or more streams when they wish to consume or just distribute their content. To participate in a stream a peer must discover and connect to other peers in the swarm carrying it and exchange chunks of content. Knowledge about other peers is obtained from a distributed set of per-stream LTs and knowledge about the LT per stream is maintained by the DOAT system. The following paragraphs illustrate the interactions between our system components and the actions necessary for a peer to receive swarmed content (see Figure 6).

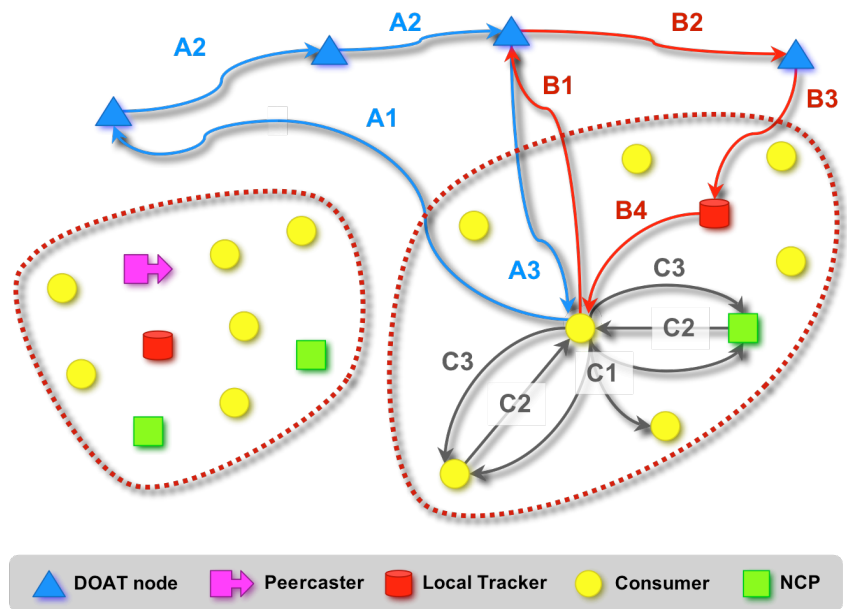

**Figure 6 – Peer interactions overview**

When a peer wishes to join the system and subsequently participate in particular streams its first task is to discover its closest DOAT node. The peer must first position itself in delay space by performing a series of delay measurements to a set of locations, as in Vivaldi [4], and use these to calculate its synthetic coordinates. Armed with these coordinates the peer queries any known node on the DOAT overlay for the closest DOAT node in the synthetic coordinate space (A1, Figure 6). The DOAT node receiving the query forwards it to the node closest in delay space to the querying peer (A2) and the identifier of the nearest DOAT node is returned (A3). This is the point of contact for all subsequent queries made by that peer, and, because it is the closest it is the best in terms of minimising stream start-up delay for that peer.

Before joining a particular stream a peer must first find its LT. It is assumed that peers discover the identifiers of the streams they wish to view through an out-of-band process (e.g. webpage hyperlinks, electronic programme guide or gossiped through a social network). The peer will use the stream identifier to query the DOAT node identified in the previous steps for the closest LT for that stream (B1). The DOAT system will then route the query to the closest LT, as described in Section 4. The query is first routed to the DOAT node closest to LT closest to the peer (B2) who subsequently forwards the request to the LT (B3). The LT then registers the peer in its database, together with its coordinates and other attributes such as upload capacity, and returns the list of registered peers back to the peer who originated the query (B4). At this point the peer has a list of all of the peers participating in the desired stream in the local area together with their attributes such as coordinates in delay space and their capacity.

Given the list of peers participating in the stream and their attributes the peer can determine to which peers it should connect according to the principles of real-time swarming over a QoS-based overlay as outlined in Sections 2 and 3. Peers issue requests for chunks to other peers (C1). The chunks are transmitted to the requesting peer (C2) and payments are made (C3). Payments are shown in the figure as being passed in the direction of receiver to sender although this is only a logical transfer as trust is the basis of our incentives mechanism and the establishment of trust following any transaction happens in the opposite direction: the receiver has more trust in the sender after a chunk has been received.

Figure 6 also shows an example of an NCP supplying the peer with chunks. The NCP receives payment (actually the NCP builds trust in the peer receiving the chunk) which is stored and can be used in other streams at other times, thanks to the trust shifting mechanism described in Section 6, to pay for chunks in those other streams and therefore to improve experienced QoS. Although the peer acting in the NCP role is not benefiting directly by consuming the content of the stream it is incentivised to participate in the local area shown in Figure 6 increasing the capacity of the overlay for that stream and reducing overall playout lag for the peers. This is an example of the incentives mechanism working in practice.

## 8 Conclusions

This article has discussed the problems of delivering real-time content over the Internet and has outlined a novel approach for swarming such content over peer-to-peer networks in a scalable manner without resorting to centralised resources. Synthetic network coordinates model the position of peers in delay space and are used to assist peers in building a QoS-aware overlay network and to schedule chunk requests and transfers over the overlay so as to reduce playout lag and increase playout continuity. Scalability and resilience is enhanced by the use of distributed local trackers, which are, in turn, managed by a novel Distributed Overlay Anycast Table. The design of the DOAT overlay and the use of Local Trackers is to ensure that queries are resolved quickly so that start-up delay is kept small. Non-Consuming Peers are encouraged to participate in streams by use of an incentives mechanism that allows contributions by a peer to be rewarded in other streams at different times. This is achieved in an entirely distributed manner without requiring a central currency repository. We have demonstrated how NCPs not only increase the available bandwidth to a swarm but also reduce playout lag.

## 9 References

[1] V. Venkataraman, K. Yoshida, and P. Francis, "Chunkyspread: Heterogeneous Unstructured Tree-Based Peer-to-Peer Multicast," in Proc. IEEE Int. Conf. Network Protocols, K. Yoshida, Ed. Santa Barbara, CA, USA 2006, pp. 2-11.

[2] B. Cohen. Incentives Build Robustness in BitTorrent. In Proc. of the Workshop on Economics of Peer-to-Peer Systems (P2PEcon'03), Berkeley, CA, June 2003.

[3] X. Hei, C. Liang, J. Liang, Y. Liu, and K. W. Ross, "A Measurement Study of a Large-Scale P2P IPTV System", in IEEE Trans. Multimedia, vol. 9, pp. 1672-1687, Dec. 2007.

[4] Dabek, F., Cox, R., Kaashoek, F., and Morris, R. 2004. Vivaldi: a decentralized network coordinate system. Proceedings of the 2004 Conference on Applications, Technologies, Architectures, and Protocols For Computer Communications (Portland, Oregon, USA,

2004). SIGCOMM '04. ACM, New York, 15-26.

[5] Stoica, I., Morris, R., Karger, D., Kaashoek, M. F., and Balakrishnan, H. 2001. Chord: A scalable peer-to-peer lookup service for internet applications. Proceedings of the 2001 Conference on Applications, Technologies, Architectures, and Protocols For Computer Communications (San Diego, California, United States). SIGCOMM '01. ACM, New York, NY, 149-160.

[6] Rolf Niedermeier, Klaus Reinhardt, Peter Sanders, Towards optimal locality in mesh-indexings. Discrete Applied Mathematics, Volume 117, Issues 1-3, , 15 March 2002, Pages 211-237.

[7] B. H. Bloom. Space/time trade-offs in hash coding with allowable errors. ACM Commun. Mag., 13(7):422-426, 1970.

[8] Hughes, D., Coulson, G., and Walkerdine, J. 2005. Free Riding on Gnutella Revisited: The Bell Tolls?. IEEE Distributed Systems Online 6, 6 (Jun. 2005), 1.

[9] Resnick, P., Kuwabara, K., Zeckhauser, R., and Friedman, E. 2000. Reputation systems. Commun. ACM 43, 12 (Dec. 2000), 45-48.

[10] A. V. Goldberg and R. E. Tarjan. A new approach to the maximum-flow problem. J. ACM, 35(4):921–940, 1988.